\documentclass[onecolumn,draftcls, 12pt]{IEEEtran}

\ifCLASSINFOpdf
  \usepackage[pdftex]{graphicx}
\else
  \usepackage[dvips]{graphicx}
\fi
\usepackage{epsfig, eucal, amssymb}
\usepackage[cmex10]{amsmath}
\usepackage{subfigure}

\begin{document}
\newtheorem{prop}{Proposition}
\newtheorem{theorem}{Theorem}
\newtheorem{lemma}{Lemma}
\newtheorem{cor}{Corollary}
\newtheorem{definition}{Definition}
\newtheorem{remark}{Remark}
\newtheorem{ach}{Achievability}
\newtheorem{con}{Converse}
%
\title{On the Multiplexing Gain of\\
$K$-user Partially Connected Interference Channel}

\author{Sang Won Choi, \IEEEmembership{Student Member, IEEE}, and Sae-Young Chung, \IEEEmembership{Senior Member, IEEE}}


\maketitle
\begin{abstract}
The multiplexing gain (MUXG) of $K$-user interference
channel (IC) with partially connected interfering links is analyzed. The
motivation for the partially connected IC comes from the fact that
not all interferences are equally strong in practice. The MUXG is characterized
as a function of the number ($K$)
of users and the number ($N \geq 1$) of interfering links. Our analysis is
mainly based on the interference alignment (IA) technique to mitigate
interference. Our main results are as follows:
One may expect that higher MUXG can be attained when
some of interfering links do not exist.
However, when $N$ is odd and $K=N+2$, the MUXG is not increased beyond the optimal
MUXG of fully connected IC, which is $\frac{KM}{2}$.
The number of interfering
links has no influence on the achievable MUXG using IA, but
affects the efficiency in terms of the number of required channel
realizations: When $N=1$ or $2$, the optimal MUXG of the fully connected IC is achievable with a finite number of channel
realizations. In case of $N \geq 3$, however, the MUXG of $\frac{KM}{2}$ can be achieved asymptotically
as the number of channel realizations tends to infinity.
\end{abstract}
\section{Introduction}
The interest for an efficient communication with limited resources
(for example, power, time, and frequency) has led us to a cellular network
system. Due to the resource sharing, however, interference is
unavoidable in the cellular network, which results in the degradation
of the throughput.
When we shift from point-to-point channel to multipoint-to-multipoint channel, in addition,
the interference management is one of the key issues for
maximizing system throughput. Thus,
the information theoretic understanding of the interference channel
(IC) motivates our interest as a basic building block for the
shared communication situation.

The asymptotic throughput performance of the IC has been studied recently \cite{Jafar07, Cadambe107}.
Specifically, in case of $2$-user multiple-input multiple-output (MIMO) Gaussian IC with
an arbitrary number of antennas, the zero forcing (ZF) has been shown to achieve the upper-bound (UB) on the
multiplexing gain (MUXG), which is derived by reducing the noise covariances and using MIMO strong IC results
\cite{Jafar07}. When the number of users
is greater than or equal to $3$, however, the ZF has limitation in mitigating interference effectively.
As a solution to this limitation, a new interference mitigation scheme called `interference alignment (IA)' has been
proposed \cite{Cadambe107} for $K$-user IC: The optimal MUXG of $\frac{K}{2}$ has been shown to be achieved
provided that many different channel realizations are available. Main idea is that the interference is
suppressed by choosing transmit beamforming vectors to align
some interferences at the desired receiver such that interferences
do not swamp all the degrees of freedom at the receiver. Recently,
the IA which requires only local channel knowledge via iterative algorithms
has been proposed in \cite{Gomadam08}.

In this correspondence, we consider the $K$-user IC with an arbitrary number of
interfering links, i.e., partially connected IC and analyze the achievable MUXG using the IA. The
motivation for the partially connected IC comes from the fact that
the interfering link far away from a receiver is negligible in the practical sense.
Depending on the path loss exponent that affects signal attenuation
as a function of distance, the effective number of interfering
transmitters can vary. The main contribution of this correspondence is as follows:
First, we derive UBs and lower-bounds (LBs) on the MUXG according to the number of
users and the number of interfering links. Second, we examine
when the LB coincides with the UB.

The structure of this correspondence is as follows: In Section II, we
describe the channel model. Along with the comment on IA in Section III, the MUXG of $K$-user IC with an arbitrary number of
interfering links is analyzed in Section IV. Finally, we conclude
our correspondence in Section V.

\section{Channel model}
\begin{figure}[!t]
\centering
\includegraphics[width=2.2in]{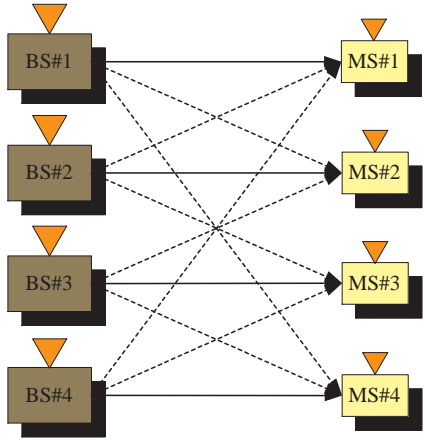}
\caption{Interference channel with $K=4$ and $N=2$ (solid arrow: desired link, dotted arrow: interfering link)}
\label{wynermodel}
\end{figure}

In this section, we describe our partially connected $K$-user IC
model with $K$ transmitters and $K$ receivers. Each transmitter and
receiver are equipped with $M~(\geq 1)$ antennas. We assume each
receiver gets interference from $N~(\leq K-1)$ adjacent transmitters
in a cyclic way. Specifically, when we arrange the indices of $K$
transmitters cyclically, the $k$-th receiver gets interference from
preceding $\lceil \frac{N}{2} \rceil$ transmitters and following
$\lfloor \frac{N}{2} \rfloor$ transmitters, where $\lceil
x \rceil$ ($\lfloor x \rfloor$) is the smallest
(biggest) integer bigger (smaller) than or equal to
$x$. Fig. $\ref{wynermodel}$ describes the $K$-user IC with $K=4$ and
$N=2$, where each mobile station (MS) gets interference from
$N$ adjacent base stations (BSs).
Note that $N=K-1$ means the fully connected $K$-user IC.

In this channel environment, the received signal vector ${\bf
Y}^{[j]}$ of the $j$-th receiver is represented as
\begin{align}
{\bf Y}^{[j]}=\sum_{i=1}^K {\bf H}^{[j,i]} {\bf X}^{[i]}+{\bf Z}^{[j]}, \quad j=1,~2,~\cdots,~K
 \label{channel_model}
\end{align}
at a specific time and frequency slot. Here, the $(r,t)$-th element
of the $M \times M$ channel matrix ${\bf H}^{[j,i]}$ represents the
channel coefficient from the $t$-th antenna of the $i$-th
transmitter to the $r$-th antenna of the $j$-th receiver, and ${\bf
X}^{[i]}$ is the signal vector of the $i$-th transmitter. Here, the channel
matrix changes independently in each slot. The noise
vector ${\bf Z}^{[j]}$ has $M$ noise elements which are independent
and identically distributed (i.i.d.) complex Gaussian with zero mean
and unit variance. The noise components are assumed to be independent
of all the transmitted signals.

When $N < K-1$,
we have $K$-user IC free of some
interfering links. For example, when $N=1$, the following matrices
are zero matrices for $j=1,~2,~\cdots,~K$.
\begin{align}
{\bf H}^{[j,i]} =0, \quad \mbox{unless}~i=j~\mbox{or}~i={\left\{(j-2) \!\!\!\! \mod K \right\} +1}.
\end{align}
The following shows the main assumptions in this correspondence.
Channel matrix assumption is applied to interfering links and desired links.
\begin{enumerate}
\item Signals are encoded over multiple time slots or multiple
frequency slots.
\item All elements of channel matrices ${\bf
H}^{[j,i]}$'s are drawn i.i.d. from a continuous channel distribution.
\item The channel is assumed to be block fading, i.e., the channel
state is fixed within a slot and changes independently from slot to slot.
\item Channel state information (CSI) is known in advance at all nodes.
\begin{enumerate}
\item When encoding is accomplished over multiple time slots, all
nodes know CSI noncausally as was done in \cite{Cadambe107}. Note
that noncausal CSI does not exist. However, one can wait for desired
CSI states and transmit partial codewords incrementally if delay is not a problem.
This is equivalent to having noncausal CSI.
\item When we encode signals over multiple frequency slots, all CSI states over multiple frequency slots are known at all nodes.
Note that coding over frequency slots is better for implementations than coding over time slots.
\end{enumerate}
\end{enumerate}

Let $n_t$ and $n_f$ denote the numbers of time and frequency slots
over which signals are encoded, respectively.
We define ${\bar{\bf H}}^{[j,i]}$ $(i,~j=1,~2,~\cdots,~K)$ as a block diagonal matrix representing
the channels from the $i$-th transmitter to the $j$-th receiver, where
the $(l,l)$-th block is a full-rank channel matrix of size $M \times M$ for $l=1,~2,~\cdots,n_t n_f$.
Note that the block diagonal matrices ${\bar{\bf H}}^{[j,i]}$'s ($i,~j=1,~2,~\cdots,~K$) are of size $M n_t n_f \times
M n_t n_f$.

\section{Interference alignment}
When we deal with multipoint-to-multipoint channel, interference management is one of the key
issues for MUXG. In the sense of suppressing interference, the IA scheme is very effective \cite{Cadambe107}.
In case of $K$-user SISO IC, the IA uses `supersymbol' coded over multiple time slots or frequency slots.

Assuming that both transmitters and receivers know all the channel
coefficients over multiple time slots or frequency slots, we have a structured MIMO
with zero off-diagonal terms. Then, we attain MUXG proportional to the number of users based on the IA scheme.
This is possible by choosing transmit beamforming vectors such that interferences from other users
are aligned to minimize the degrees of freedom occupied by the subspace from the interferences at each receiver.
The IA is feasible because there are multiple $(n_t n_f)$ different realizations of channels.

The definition of MUXG \cite{Zheng03} in $K$-user IC is
\begin{align}
\Gamma=\underset{\mbox{\small SNR} \rightarrow \infty}{\lim}
&\frac{C_{+}(\mbox{SNR})}{\log(\mbox{SNR})},
\end{align}
where $C_{+} (\mbox{SNR})$ is the sum capacity (per slot) at signal-to-noise ratio (SNR). Note that the ZF is enough to achieve the optimal multiplexing
gain for the point to point channel, MAC, and
BC \cite{Jafar07}.

In case of fully connected IC with $K \geq 3$, along with ZF at the
receiver, the transmit beamforming such that other interferences are
aligned is required to achieve the optimal MUXG of $\frac{K}{2}$ asymptotically when $M=1$. When $K=3$ with $M \geq 2$,
it is shown that we achieve the optimal MUXG of $\frac{KM}{2}$ with only one slot, i.e., $n_t n_f=1$ \cite{Cadambe107}.
\section{Multiplexing gain}
\subsection{Upper-bound}
When $K=2$, we have the following UB on the MUXG assuming that all the channel matrices are full-rank. It is obtained by reducing noise covariance at each receiver
such that the channel matrices of the interfering links are diagonalized and strong
interference conditions are met. Then, we have two resulting
MIMO MACs and get the UB on the MUXG as the minimum of MUXGs of the two MACs.
Explicit expression is as follows:
\begin{lemma} The optimal MUXG of
$2$-user Gaussian IC comprised of $M_i$ transmit and $N_i$ receive antennas
for user $i$ ($i=1,~2$) is
\begin{eqnarray}
\Gamma&=&
\min\left\{M_1+M_2,~N_1+N_2,~\max(M_1,N_2),~\max(M_2,N_1)\right\}.
\label{2user_outerbound}
\end{eqnarray}
\label{outer_lemma}
\end{lemma}
\begin{proof}
It follows from {\em Theorem $2$} in \cite{Jafar07}.
\end{proof}
\begin{theorem}
When we have $M_i$ transmit and $N_i$ receive antennas for user
$i$ $(i=1,~2,~\cdots,~K)$, the MUXG $\Gamma$ is upper-bounded by
\begin{align}
\Gamma \leq& \frac{1}{K-1} \sum_{k=1}^{\frac{K(K-1)}{2}} \Gamma_k,
\label{outerbound}
\end{align}
where $\Gamma_k$ is
\begin{align*}
&\min\left\{M_{\pi(k,1)}+M_{\pi(k,2)},~N_{\pi(k,1)}+N_{\pi(k,2)},~\max(M_{\pi(k,1)},N_{\pi(k,2)}),
~\max(M_{\pi(k,2)},~N_{\pi(k,1)})\right\}.
\end{align*}
Here, $\pi(k,i)~(i=1,~2)$ is the $i$-th component of the $k$-th
combination among $_K C _2$ sets comprised of unordered two user indices.
$(1,2),~(1,3),~\cdots,~(K-1,K)$. For example, when $K=3$, we have
$_3C_2=3$ index sets $\{1,2\},~\{1,3\},~\{2,3\}$ and
$\pi(1,1)=1$, $\pi(1,2)=2$, $\pi(2,1)=1$, $\pi(2,2)=3$, $\pi(3,1)=2$, and $\pi(3,2)=3$.
\label{outerbound_proof}
\end{theorem}
\begin{proof}
Without loss of generality, we pick nodes $1$ and $2$, respectively.
Then, the other transmit nodes $3,~4,~\cdots,~K$ interfere with
receive nodes $1$ and $2$.
Then, we obtain the UB from $(\ref{2user_outerbound})$
by ignoring such interferences.

In a similar manner, we have the UB on the MUXG for other
user pairs. Since there are $\frac{K(K-1)}{2}$ user pairs resulting in $K-1$ times the sum of the individual rates, we get
$(\ref{outerbound})$.
\end{proof}
\begin{cor}
The optimal MUXG of $K$-user fully connected IC with $M$ antennas at each
nodes is less than or equal to $\frac{KM}{2}$.
\label{outerMUXG}
\end{cor}
\begin{proof}
It is straightforwardly obtained from {\em{Theorem $\ref{outerbound_proof}$}} when we use $M_i=N_j=M~(i,~j=1,~2,~\cdots,~K)$.
\end{proof}

\begin{remark}
From {\em Corollary} $\ref{outerMUXG}$ with $M=1$,
the UB of $\frac{K}{2}$ \cite{Madsen05} on the MUXG is derived in the $K$-user SISO IC.
\end{remark}


\begin{cor}
In partially connected $K$-user ICs,
the UB of $\frac{KM}{2}$ on the MUXG is still maintained only when
we have $N=2p+1$ and $N=K-2$ with a nonnegative integer $p$.
\label{strange_OB}
\end{cor}

\begin{proof}
From {\em Theorem $\ref{outerbound_proof}$}, it is clear that the UB of $\frac{KM}{2}$
on the MUXG is maintained if and only if any $2$-user indices in the user index set $\{1,~2,~\cdots,~K\}$
form a $2$-user IC with at least one interfering link. Let the number of
such $2$-user ICs be $T$. Since $T=\frac{K(K-1)}{2}$ is equivalent to the UB of $\frac{KM}{2}$ on the MUXG from
{\em Theorem} $\ref{outerbound_proof}$, it suffices to find minimum $N$ which guarantees
\begin{eqnarray}
T=\frac{K(K-1)}{2}.
\label{outerN1}
\end{eqnarray}
Note that $N$ and $N+1$ result in same $T$ for odd $N$.
\begin{itemize}
\item When $K$ is even, we obtain $T=Kp+\frac{K}{2}$ for $N=2p+1$.
From $(\ref{outerN1})$, we have $N=K-1$, which means that only fully connected $K$-user IC
guarantees the outer bound of $\frac{KM}{2}$ on the MUXG.
\item When $K$ is odd, we have $T=K(p+1)$ for $N=2p+1$. From $(\ref{outerN1})$, we get
$N=K-2$. Thus, partially connected $K$-user IC with $N=K-2$ has
the outer bound of $\frac{KM}{2}$ on the MUXG.
\end{itemize}
\end{proof}

\begin{remark}
Except for the cases in {\em Corollary} $\ref{strange_OB}$,
the UBs on the MUXGs for $K$-user partially connected ICs are
always greater than $\frac{KM}{2}$.
\end{remark}



\subsection{Lower-bound when $N=1$}
Now, we consider the achievable MUXG with respect to the number ($N$) of interfering links.
Note that the number ($M$) of antennas and the number ($K$) of users are arbitrary.

\begin{theorem}
When $N=1$ in $K$-user IC, the MUXG of $\frac{KM}{2}$
is achieved based on simple time division multiplexing (TDM) or ZF with number of required slots (NRS) of $1$ or $2$.
\label{achievability_N1}
\end{theorem}
\begin{proof}
We prove this for all cases of $M$ and $K$.
\begin{itemize}
\item $M=1$: The TDM strategy with NRS of $1$ gives us the MUXG of $\frac{K}{2}$ and $\frac{K-1}{2}$ for even $K$ and odd $K$, respectively.
When the ZF is used with NRS of $2$, the MUXG of $\frac{K}{2}$ is achieved for all $K$.
Note that in case of fully connected $K$-user IC with $M=1$ and $K \geq 3$, the NRS to achieve the MUXG of $\frac{K}{2}$ goes to infinity \cite{Cadambe107}.

\item $M$ and $K$ are even:
We obtain MUXG of $\frac{KM}{2}$ by the TDM strategy.
Alternatively, the MUXG of $\frac{KM}{2}$ is achievable by the ZF.
One thing worth noting is that the NRS is only $1$ in both cases.
\item $M$ is even and $K$ is odd:
The MUXG of $\frac{(K-1)M}{2}$ with NRS of $1$ is achievable by alternating between even- and odd-indexed users.
We can increase the achievable MUXG by the following procedure: Each transmitter transmits $\frac{M}{2}$ streams, each receiver gets the desired streams
by ZF. Then, we have the MUXG of $\frac{KM}{2}$ with NRS of $1$.
\item $M \geq 3$ is odd and $K$ is even:
From the TDM strategy, the MUXG of $\frac{KM}{2}$ is achievable with NRS of $1$. One interesting thing is that
the ZF gives us the MUXG of $\frac{KM}{2}$, where even- and odd-indexed users transmit $\frac{M+1}{2}$ and $\frac{M-1}{2}$ streams, respectively. At each receiver, the ZF is used to extract
the desired streams. Then, we attain the MUXG of
\begin{eqnarray*}
\frac{K}{2} \cdot \frac{M+1}{2} + \frac{K}{2} \cdot \frac{M-1}{2} = \frac{KM}{2}
\end{eqnarray*}
with NRS of $1$.
\item $M \geq 3$ and $K$ are odd:
we have the following achievable MUXG with respect to NRS.
First, when we use TDM strategy, we have MUXG of $\frac{(K-1)M}{2}$ with NRS of $1$. Second,
the MUXG of $\frac{(KM-1)}{2}$ is attained by ZF,
where each of even- and odd-indexed users transmits $\frac{M+1}{2}$ and $\frac{M-1}{2}$ streams,
respectively. Then, each receiver gets the desired streams by ZF. Thus, we achieve the MUXG of
\begin{eqnarray*}
\frac{K-1}{2} \cdot \frac{M+1}{2} + \frac{K+1}{2} \cdot \frac{M-1}{2} = \frac{KM-1}{2}
\end{eqnarray*}
with NRS of $1$.
Another way to achieve the MUXG of $\frac{KM}{2}$ is that all transmitters transmit $M$ streams over $2$ slots and all the receivers get the
desired streams by ZF.
\end{itemize}
\end{proof}

\begin{remark}
Since the number of interferences is $1$ per each receiver, the IA is not applicable, and the
ZF at each receiver is
sufficient to mitigate the adjacent interference effectively in the
sense of the MUXG.
\end{remark}

In Table \ref{MUXgaintableN1}, we summarize the MUXG with respect to NRS when $N$$=$$1$ or $2$. Note that
the IA implicitly includes ZF when $N$$=$$1$, which means that the IA becomes the ZF when $N=1$. It is seen that
the MUXG of TDM is comparable to that of ZF. Since the ZF requires relatively more CSI than the TDM, the TDM strategy
seems to be more reasonable in the sense of the MUXG per required CSI overhead.
However, when $M$ is large enough with odd $K$, the loss in the MUXG
becomes in no way negligible.

\begin{remark}
When the number ($K$) of users tends to infinity,
the MUXG from TDM strategy is asymptotically same as that from ZF.
\end{remark}

\begin{remark}
For all cases of $M$ with $K=3$, ZF gives us optimal MUXG with
NRS of $1$ or $2$.
\end{remark}
\begin{remark}
In case of odd $M$ and odd $K$, NRS of $1$ with ZF is
enough to approach MUXG of $\frac{KM}{2}$ when either $K$ or $M$ tends to infinity.
\end{remark}

\begin{remark}
It seems to be reasonable to use TDM strategy when $K$ is large enough and
ZF when either $K$ or $M$ is large enough.
\end{remark}
\subsection{Lower-bound when $N=2$}
\begin{theorem}
When $N=2$ in $K$-user IC, the MUXG of $\frac{KM}{2}$
is available with NRS of $1$ or $2$.
\label{inner_N2}
\end{theorem}
\begin{proof}
Assume that $M$ is even.
The IA conditions are formed by aligning interferences at each receiver and
we have the following IA conditions
\begin{eqnarray*}
{\bar{\bf H}}^{[1,K]} {\bf V}^{[K]} &\doteqdot& {\bar{\bf H}}^{[1,2]} {\bf V}^{[2]},\cr
{\bar{\bf H}}^{[2,1]} {\bf V}^{[1]} &\doteqdot& {\bar{\bf H}}^{[2,3]} {\bf V}^{[3]},\cr
{\bar{\bf H}}^{[3,2]} {\bf V}^{[2]} &\doteqdot& {\bar{\bf H}}^{[3,4]} {\bf V}^{[4]},\cr
&\vdots&\cr
{\bar{\bf H}}^{[K-2,K-3]} {\bf V}^{[K-3]} &\doteqdot& {\bar{\bf H}}^{[K-2, K-1]} {\bf V}^{[K-1]}, \cr
{\bar{\bf H}}^{[K-1, K-2]} {\bf V}^{[K-2]} &\doteqdot& {\bar{\bf H}}^{[K-1, K]} {\bf V}^{[K]},
\end{eqnarray*}
and
\begin{eqnarray}
{\bar{\bf H}}^{[K, K-1]} {\bf V}^{[K-1]} &\doteqdot& {\bar{\bf H}}^{[K, 1]} {\bf V}^{[1]},
\label{N2_achievability_odd}
\end{eqnarray}
where ${\bf V}^{[k]}$ is the $M \times \frac{M}{2}$ transmit beamforming matrix of the
$k$-th user $(k=1,~2,~\cdots,~K)$ and $P \doteqdot Q$ means that the column space of $P$ is the same as that of $Q$.
Here, the first equation in $(\ref{N2_achievability_odd})$ comes from
the condition that the interferences from adjacent $2$-nd and $K$-th transmitters
with beamforming vectors in ${\bf V}^{[2]}$ and ${\bf V}^{[K]}$
are aligned at receiver $1$. The remaining conditions are obtained in a similar manner.

When $K$ is odd, the transmit beamforming matrices
${\bf V}^{[k]}$'s ($k=1,~2,~\cdots,~K$) are obtained in the following
order
\begin{eqnarray}
&&{\bf V}^{[2]} \rightarrow {\bf V}^{[4]} \rightarrow \cdots
\rightarrow {\bf V}^{[K-1]} \rightarrow {\bf V}^{[1]}
\rightarrow {\bf V}^{[3]} \cdots \rightarrow {\bf V}^{[K]} \rightarrow {\bf V}^{[2]}
\label{beamform_order_odd}
\end{eqnarray}
from (\ref{N2_achievability_odd}). Once we set the transmit beamforming matrix ${\bf V}^{[2]}$, we
obtain ${\bf V}^{[4]}$ and ${\bf V}^{[K]}$ from ${\bf V}^{[2]}$ using $(\ref{N2_achievability_odd})$. Next, ${\bf V}^{[6]}$ and ${\bf V}^{[K-2]}$
are derived from the known values ${\bf V}^{[4]}$ and ${\bf V}^{[K]}$ using $(\ref{N2_achievability_odd})$.
All the remaining transmit beamforming matrices are also obtained by the IA conditions $(\ref{N2_achievability_odd})$.

In $(\ref{beamform_order_odd})$, an initial transmit beamforming matrix ${\bf V}^{[2]}$ is composed of any set of
$\frac{M}{2}$ eigenvectors of
\begin{eqnarray*}
A&=& \left\{({\bar{\bf H}}^{[3,2]})^{-1} {\bar{\bf H}}^{[3,4]} ({\bar{\bf H}}^{[5,4]})^{-1} {\bar{\bf H}}^{[5,6]}  \cdots \right. \cr
&&\left.({\bar{\bf H}}^{[K-2,K-3]})^{-1} {\bar{\bf H}}^{[K-2,K-1]}  \right\} \cdot ({\bar{\bf H}}^{[K,K-1]})^{-1} {\bar{\bf H}}^{[K,1]} \cr
&&\cdot \left\{({\bar{\bf H}}^{[2,1]})^{-1} {\bar{\bf H}}^{[2,3]} ({\bar{\bf H}}^{[4,3]})^{-1} {\bar{\bf H}}^{[4,5]} \cdots \right. \cr
&& \left. ({\bar{\bf H}}^{[K-1,K-2]})^{-1} {\bar{\bf H}}^{[K-1,K]}  \right\} \cdot ({\bar{\bf H}}^{[1,K]})^{-1} {\bar{\bf H}}^{[1,2]},
\end{eqnarray*}
which is full-rank with probability $1$.

Finally, each $k$-th transmitter $(k=1,~2,~\cdots,~K)$ transmits $\frac{M}{2}$ streams with transmit beamforming matrix
${\bf V}^{[k]}$ and the corresponding receiver $k$ obtains the $\frac{M}{2}$ streams per slot by ZF.

Second, when $K$ is even, the transmit beamforming matrices ${\bf V}^{[k]}$'s $(k=1,~2,~\cdots,~K)$ are chosen in the following order
\begin{eqnarray*}
&&{\bf V}^{[1]} \rightarrow {\bf V}^{[3]} \rightarrow \cdots \rightarrow {\bf V}^{[K-3]} \rightarrow {\bf V}^{[K-1]}
\end{eqnarray*}
and
\begin{eqnarray*}
&&{\bf V}^{[2]} \rightarrow {\bf V}^{[4]} \rightarrow \cdots \rightarrow {\bf V}^{[K-2]} \rightarrow {\bf V}^{[K]},
\end{eqnarray*}
where two initial transmit beamforming matrices ${\bf V}^{[1]}$ and ${\bf V}^{[2]}$ consist of any set of $\frac{M}{2}$
eigenvectors of
\begin{eqnarray*}
B&=& \left\{ ({\bar{\bf H}}^{[2,1]})^{-1} {\bar{\bf H}}^{[2,3]} ({\bar{\bf H}}^{[4,3]})^{-1} {\bar{\bf H}}^{[4,5]} \cdots \right. \cr
&&\left. ({\bar{\bf H}}^{[K-2,K-3]})^{-1} {\bar{\bf H}}^{[K-2,K-1]} \right\} \cr
&& \cdot ({\bar{\bf H}}^{[K,K-1]})^{-1} {\bar{\bf H}}^{[K,1]}
\end{eqnarray*}
and
\begin{eqnarray*}
C&=& \left\{ ({\bar{\bf H}}^{[3,2]})^{-1} {\bar{\bf H}}^{[3,4]} ({\bar{\bf H}}^{[5,4]})^{-1} {\bar{\bf H}}^{[5,6]} \cdots \right. \cr
&& \left. ({\bar{\bf H}}^{[K-1,K-2]})^{-1} {\bar{\bf H}}^{[K-1,K]} \right\} \cr
&& \cdot ({\bar{\bf H}}^{[1,K]})^{-1} {\bar{\bf H}}^{[1,2]},
\end{eqnarray*}
respectively.
Thus, interference-free $\frac{KM}{2}$ streams per slot
are obtained by ZF at each receiver, which gives us MUXG of $\frac{KM}{2}$.

When $M$ is odd, we have essentially the same IA conditions $(\ref{N2_achievability_odd})$ by using $2$ slots. Even though the form
of the channel matrix is block diagonalized, the fact that the channel matrices
$A,~B$, and $C$ are full-rank is not changed.
Thus, the MUXG of $\frac{KM}{2}$ is also obtained with $2$ slots in case of odd $M$. Another way is that even- and
odd-indexed users transmit different amount of streams:
When $K$ is even or odd, the transmit beamforming matrices ${\bf V}^{[k]}$'s $(k=1~,2,~\cdots,~K)$ are constructed from the IA conditions
$(\ref{N2_achievability_odd})$. Note that the size of the transmit beamforming matrices are $M \times \frac{M+1}{2}$. Even-indexed users
transmit $\frac{M+1}{2}$ streams, and odd-indexed users transmit $\frac{M-1}{2}$ streams.
By extracting the desired streams using ZF, we have
\begin{eqnarray}
\frac{K}{2} \cdot \frac{M+1}{2} + \frac{K}{2} \cdot \frac{M-1}{2} &=& \frac{KM}{2}
\label{asymIA1}
\end{eqnarray}
and
\begin{eqnarray}
\frac{K-1}{2} \cdot \frac{M+1}{2} + \frac{K+1}{2} \cdot \frac{M-1}{2} &=& \frac{KM-1}{2}
\label{asymIA2}
\end{eqnarray}
when $K$ is even and odd, respectively.
\end{proof}

\begin{remark}
When $K$ and $M$ are odd and either $K$ or $M$ tends to infinity, the MUXG of $\frac{KM}{2}$ is achieved asymptotically
by IA scheme with NRS of $1$.
\end{remark}

\begin{remark}
When $N=2$, the ZF itself at each receiver is not enough to achieve the optimal MUXG since the number of interferences to each receiver
is greater than $1$ and interferences need to be aligned to minimize the dimension of the signal space
occupied by the interferences.
\end{remark}

As shown in Table \ref{MUXgaintableN1}, the MUXG of $\frac{KM}{2}$
is still achievable by the IA when the number ($N)$ increases from
$1$ to $2$. Considering the CSI overheads, the TDM rather
than the IA seems to be reasonable since the IA requires all the CSI states
of all links in general. All the MUXG characteristics of the TDM when $N=2$ are exactly same as those when $N=1$.
\begin{table}[!t]
\caption{MUXG with respect to NRS when $N$$=$$1$ or $2$}
\label{MUXgaintableN1}
\renewcommand{\arraystretch}{1.5}
\begin{center}
\begin{tabular}{||c|c|c|c|c|c||}
\hline
\cline{4-6}
\multicolumn{3}{||c|}{}& {TDM} & \multicolumn{2}{|c||}{IA}\\
\hline
{}& {$K$} & {MUXG} & {$\frac{KM}{2}$} &  \multicolumn{2}{|c||}{$\frac{KM}{2}$}\\
\cline{3-6}
{$M$}& {(even)} & {NRS} & {$1$} & \multicolumn{2}{|c||}{$1$}\\
\cline{2-6}
{(even)}& {$K$} & {MUXG} & {$\frac{(K-1)M}{2}$} & \multicolumn{2}{|c||}{$\frac{KM}{2}$}\\
\cline{3-6}
{}& {(odd)} & {NRS} & {$1$} & \multicolumn{2}{|c||}{$1$}\\
\hline
{}& {$K$} & {MUXG} & {$\frac{KM}{2}$} & \multicolumn{2}{|c||}{$\frac{KM}{2}$}\\
\cline{3-6}
{$M \geq 3$}& {(even)} & {NRS} & {$1$} & \multicolumn{2}{|c||}{$1$}\\
\cline{2-6}
{(odd)}& {$K$} & {MUXG} & {$\frac{(K-1)M}{2}$} & {$\frac{KM-1}{2}$} & {$\frac{KM}{2}$}\\
\cline{3-6}
{}& {(odd)} & {NRS} & {$1$} & {$1$} & {$2$}\\
\hline
\end{tabular}
\end{center}
\end{table}
\subsection{Lower-bound when $3 \leq N \leq K-1$}
\begin{lemma}
The optimal MUXG of $\frac{K}{2}$ is achieved
asymptotically with infinitely many slots in fully connected $K$-user SISO IC.
\label{asymptotic_lemma}
\end{lemma}
\begin{proof}
If follows from {\em Theorem $1$} in \cite{Cadambe107}.
\end{proof}
\begin{cor}
For the $K$-user IC with $M$ multiple antennas at each nodes, the optimal
MUXG of $\frac{KM}{2}$ is
achieved asymptotically with infinitely many slots.
\label{infinite_cor}
\end{cor}
\begin{proof}
First, when IA conditions are met for $KM$-user fully connected SISO IC, the same IA conditions are also
satisfied for the $K$-user fully connected IC with $M$ antennas.
Second, the IA conditions for $K$-user partially connected IC with $M$ antennas are less restrictive than
those for the fully connected one. Thus, the MUXG of $\frac{KM}{2}$ is achievable asymptotically
with infinitely many slots in the $K$-user partially connected MIMO IC.
\end{proof}
\begin{theorem}
The MUXG of $\frac{KM}{2}$ from encoding over a finite number of slots
is not achievable with probability $1$ in the $K$-user MIMO IC.
\label{no_N3}
\end{theorem}
\begin{proof}
Since IA conditions for $N \geq 4$ are more restrictive than those for $N=3$ and include IA conditions for $N=3$, it suffices to show that
there are no explicit transmit beamforming matrices satisfying the IA conditions for $N=3$.
We prove this using contradiction. First, we consider the case
where $M$ is even. Assume that there exist transmit beamforming matrices satisfying the following IA conditions
\begin{eqnarray*}
{\bar{\bf H}}^{[1,2]} {\bf V}^{[2]} &\doteqdot& {\bar{\bf H}}^{[1, K-1]} {\bf V}^{[K-1]}\doteqdot {\bar{\bf H}}^{[1, K]} {\bf V}^{[K]}, \cr
{\bar{\bf H}}^{[2,1]} {\bf V}^{[1]} &\doteqdot& {\bar{\bf H}}^{[2, 3]} {\bf V}^{[3]}\doteqdot {\bar{\bf H}}^{[2, K]} {\bf V}^{[K]}, \cr
{\bar{\bf H}}^{[3,1]} {\bf V}^{[1]} &\doteqdot& {\bar{\bf H}}^{[3, 2]} {\bf V}^{[2]}\doteqdot {\bar{\bf H}}^{[3, 4]} {\bf V}^{[4]}, \cr
&&~~~~~~\vdots& \cr
{\bar{\bf H}}^{[K-1, K-3]} {\bf V}^{[K-3]} &\doteqdot& {\bar{\bf H}}^{[K-1, K-2]} {\bf V}^{[K-2]}\doteqdot{\bar{\bf H}}^{[K-1, K]} {\bf V}^{[K]},
\end{eqnarray*}
and
\begin{eqnarray}
{\bar{\bf H}}^{[K, K-2]} {\bf V}^{[K-2]} &\doteqdot& {\bar{\bf H}}^{[K, K-1]} {\bf V}^{[K-1]}\doteqdot {\bar{\bf H}}^{[K, 1]} {\bf V}^{[1]}.
\label{N3_nonachievability}
\end{eqnarray}
Note that the IA conditions $(\ref{N3_nonachievability})$ are necessary conditions for
achieving the MUXG of $\frac{KM}{2}$ with a finite NRS.
From $(\ref{N3_nonachievability})$, the transmit beamforming matrices
${\bf V}^{[K-1]}$ and ${\bf V}^{[K-2]}$ must satisfy
\begin{eqnarray*}
{\bf V}^{[K-1]}&\doteqdot& ({\bar{\bf H}}^{[1, K-1]})^{-1} {\bar{\bf H}}^{[1, K]} {\bf V}^{[K]}, \cr
{\bf V}^{[K-1]}&\doteqdot& ({\bar{\bf H}}^{[K,K-1]})^{-1} {\bar{\bf H}}^{[K, 1]} ({\bar{\bf H}}^{[2,1]})^{-1} {\bar{\bf H}}^{[2, K]}{\bf V}^{[K]},\cr
{\bf V}^{[K-2]}&\doteqdot& ({\bar{\bf H}}^{[K-1, K-2]})^{-1} {\bar{\bf H}}^{[K-1, K]} {\bf V}^{[K]},
\end{eqnarray*}
and
\begin{eqnarray}
{\bf V}^{[K-2]}&\doteqdot& ({\bar{\bf H}}^{[K, K-2]})^{-1} {\bar{\bf H}}^{[K, 1]} ({\bar{\bf H}}^{[2,1]})^{-1} {\bar{\bf H}}^{[2,K]}{\bf V}^{[K]}.
\label{IA_infinite}
\end{eqnarray}
Thus, the transmit beamforming matrix ${\bf V}^{[K]}$ must have the following relations
\begin{eqnarray*}
{\bf V}^{[K]}&\doteqdot& D{\bf V}^{[K]}
\end{eqnarray*}
and
\begin{eqnarray*}
{\bf V}^{[K]}&\doteqdot& E{\bf V}^{[K]},
\end{eqnarray*}
where
\begin{eqnarray*}
D&=&({\bar{\bf H}}^{[1, K]})^{-1} {\bar{\bf H}}^{[1,K-1]} ({\bar{\bf H}}^{[K, K-1]})^{-1} {\bar{\bf H}}^{[K, 1]} \cr
  && \cdot ({\bar{\bf H}}^{[2,1]})^{-1} {\bar{\bf H}}^{[2,K]}
\end{eqnarray*}
and
\begin{eqnarray*}
E&=&({\bar{\bf H}}^{[K-1, K]})^{-1} {\bar{\bf H}}^{[K-1, K-2]}  ({\bar{\bf H}}^{[K, K-2]})^{-1} {\bar{\bf H}}^{[K, 1]}\cr
  && \cdot ({\bar{\bf H}}^{[2, 1]})^{-1} {\bar{\bf H}}^{[2, K]}.
\end{eqnarray*}
However, there is no transmit beamforming matrix ${\bf V}^{[K]}$
composed of any set of $\frac{K}{2}$ eigenvectors of $D$ and $E$ simultaneously since all the
i.i.d. components of the channel matrices are assumed to be drawn from a continuous distribution.
Thus, it is a contradiction.

When $M$ is odd, the above contradiction is also shown in essentially the same manner.
\end{proof}

\begin{table}[!t]
\caption{Classification of $K$-user IC ($\square$: UB=LB=$\frac{KM}{2}$,~$\blacksquare$: UB$>$LB=$\frac{KM}{2}$)} \label{MUXgaintable}
\renewcommand{\arraystretch}{2}
\begin{center}
\begin{tabular}{||c|c|c|c|c|c|c|c|c|c|c||}
\hline \multicolumn{3}{||c}{}& \multicolumn{8}{|c||}{$K$}\\
\cline{3-11}
\multicolumn{2}{||c|}{} & NRS& $2$ & $3$ & $4$ & $5$ & $6$ & 7 & 8 & 9\\
\cline {1-11}
{}& $1$& $1$ or $2$ &\multicolumn{2}{|c|}{$\square$}& \multicolumn{6}{|c||}{${\blacksquare}$}\\
\cline {2-11}
{}& $2$& $1$ or $2$&$\times$ & $\square$ &\multicolumn{6}{|c||}{${\blacksquare}$}\\
\cline{2-11}
{}& $3$& {}&$\times$ & $\times$ & \multicolumn{2}{|c|}{${\square}$} &\multicolumn{4}{|c||}{}\\
\cline{2-2}\cline{4-7}
{$N$}&$4$& {}&$\times$ & $\times$ & $\times$ &{${\square}$}& \multicolumn{4}{|c||}{${\blacksquare}$}\\
\cline{2-2} \cline{4-9}
{}& $5$& $\infty$ &$\times$ & $\times$ & $\times$ & $\times$ &\multicolumn{2}{|c|}{${\square}$}&\multicolumn{2}{|c||}{}\\
\cline{2-2} \cline{4-9}
{}& $6$& {}&$\times$ & $\times$ & $\times$ & $\times$ &$\times$ & ${\square}$& \multicolumn{2}{|c||}{}\\
\cline{2-2} \cline{4-11}
{}& $7$& {}&$\times$ & $\times$ & $\times$ & $\times$ & $\times$ & $\times$ & \multicolumn{2}{|c||}{$\square$}\\
\hline
\end{tabular}
\end{center}
\end{table}

\subsection{MUXG characteristics}
According to the number of users and the number of interfering links, we can classify $K$-user IC
as shown in Table \ref{MUXgaintable}.
Note that `$\times$' represents the case that cannot happen since $N \leq K-1$ in Table \ref{MUXgaintable}.

Based on the IA scheme, the MUXG of $\frac{KM}{2}$ is achieved with
a finite (or asymptotically many) NRS.
The UB is derived from {\em Theorem} $\ref{outerbound_proof}$.
Then, the following relations between LBs and UBs along with NRS are
summarized.

\begin{enumerate}
\item LBs and UBs on the MUXG
    \begin{enumerate}
    \item Optimal MUXG is equal to $\frac{KM}{2}$ ($\square$).\\
    One thing worth mentioning is when we have $N=2p+1$ and $K=N+2$ with nonnegative integer $p$:
    The optimal MUXG is equal to that of fully connected IC, which is $\frac{KM}{2}$. This is confirmed by {\em Corollary} $\ref{strange_OB}$ and
    {\em Corollary} $\ref{infinite_cor}$.
    \item The MUXG of $\frac{KM}{2}$ is achievable and does not coincide with the UB in {\em Theorem} $\ref{outerbound_proof}$
    (${\blacksquare}$).\\
    In this case, the UB on the MUXG is greater than $\frac{KM}{2}$ since
    there exists at least one $2$-user pair in the user index set such that it forms a $2$-user IC with no interfering links and results
    in increase of the UB over $\frac{KM}{2}$ from {\em Theorem} $\ref{outerbound_proof}$.
    \end{enumerate}
\item Number of required slots
    \begin{enumerate}
    \item When $N=1$ or $2$, the NRS is finite to attain the MUXG of $\frac{KM}{2}$.\\
    In case of $N=1$, simple TDM or ZF gives us the achievable MUXG of
    $\frac{KM}{2}$, which is confirmed by {\em Theorem} $\ref{achievability_N1}$.
    When $N=2$, the IA is applicable since it is possible to construct beamforming matrices satisfying
    IA conditions from {\em Theorem} $\ref{inner_N2}$.
    \item When $N \geq 3$, infinitely many slots are required to achieve the MUXG of $\frac{KM}{2}$ asymptotically.\\
    From {\em Corollary} $\ref{infinite_cor}$, the NRS goes to infinity, which is supported by {\em Theorem} $\ref{no_N3}$:
    There do not exist explicit beamforming matrices satisfying IA conditions with a finite NRS.
    \end{enumerate}
\end{enumerate}

\section{Conclusion}
As an asymptotic performance measure, the MUXG of $K$-user IC was investigated in this correspondence.
One of main results is that in terms of the NRS,
it is possible to have more efficient communications for $K \geq 4$ when the number of interfering links is $1$ or $2$,
which is not seen in fully connected IC.
But, when the number of interfering links is greater than or equal to $3$, asymptotically many slots are
still necessary for achieving the optimal MUXG.
In some cases, the UB on the MUXG in {\em Theorem} $\ref{outerbound_proof}$ does not
coincide with the MUXG achieved by the IA scheme even with asymptotically many slots.
In comparison with the fully connected IC, one might expect that
the optimal MUXG would increase when the number of interfering links decreases. Counter-intuitively, it was observed that when $N=2p+1$ and $K=N+2$ with a nonnegative integer $p$,
the MUXG is equal to the optimal MUXG of the fully connected one, which is $\frac{KM}{2}$.
As a further work, either tighter UBs or higher LBs for partially connected ICs
need to be developed.

\end{document}